# Microassembly of Multi-Material and 3D Integration Enabled by Programmable and Universal High-Precision Micro-Transfer Printing

Qinhua Guo[1]†, Lizhou Yang[1]†, Yawen Gan[1], Jingyang Zhang[1], Jiajun Zhang[1], Jiahao Jiang[1], Weihan Lin[1], Kaiqi Chen[1], Chenchen Zhang[1], Yunda Wang[1,2]*

## Affiliations

[1]Smart Manufacturing Thrust, The Hong Kong University of Science and Technology (Guangzhou); Guangzhou, 511400, China.

[2]Department of Mechanical and Aerospace Engineering, The Hong Kong University of Science and Technology; Hong Kong SAR, 999077, China.

*Corresponding author. Email: ydwang@ust.hk

† These authors contributed equally to this work.

## Abstract

Micro-transfer printing is an assembly technology that enables large-scale integration of diverse materials and components from micro- to nano-scale. However, traditional micro-transfer printing technologies lack dynamic selectivity, limiting capabilities in sorting and repairing materials and components for effective yield management during large-scale manufacturing and integration processes. In this work, we introduce a dynamically programmable micro-transfer printing system utilizing a sharp phase-changing polymer and an independently addressable microheater array to modulate adhesion through localized heating. The system demonstrates dynamically programmable capabilities for selective transfer of various materials including semiconductors, polymers and metals, handling geometries from micro-scale chiplets to nanometer-thick films and micro-spheres. It also exhibits exceptional capabilities in 3D stacking and heterogeneous materials integration, significantly advancing the manufacturability of complex microsystems. As a demonstration, we successfully perform dynamically programmable transfer of microLED chips to create arbitrarily specified patterns, offering a promising solution to the challenges of mass transfer and pixel repair in microLED display manufacturing.

## Introduction

Micro-transfer printing (μTP), an essential technology for heterogeneous integration, enables precise and flexible placement of micro- and nano-scale functionalized components onto various substrates (*1-5*). This method provides significant design flexibility and material compatibility by separating the fabrication processes of different materials and components. It facilitates the creation of complex systems by integrating diverse types of devices on a single substrate, offering substantial benefits for applications in flexible electronics (*6, 7*), microLED displays (*8, 9*), optoelectronics (*10-13*), heterogeneously integrated semiconductor circuits (*14, 15*), and microsystems (*16, 17*).

Various methods of μTP have been developed for advanced electronics manufacturing, employing mechanisms such as van der Waals force (*1, 18*), laser-induced thermal effects (*19, 20*), electrostatic interactions (*21*), electromagnetic forces (*22*), and fluidic assisted assembly (*23, 24*). Among these, the use of an elastomer stamp for transferring components via kinetic control of adhesion is one of the earliest and most extensively studied methods (*25-27*). This technology based on tuning van der Waals forces, has successfully facilitated the transfer of a diverse range of materials and devices, including silicon integrated circuits (*14*), GaAs lasers (*28*), GaN LEDs (*29, 30*), micro-disk resonators (*31, 32*), and nanowire lasers (*33, 34*), enhancing wafer-level heterogeneous integration and the development of flexible electronics. However, this method is limited by its inability to dynamically select specific subsets within a component array, which affects its potential for intelligent assembly, such as for sorting and repair operations. It also faces challenges in transferring objects with non-flat surfaces (*35*), where the limited contact area reduces adhesion. Alternatives like stamps driven by electrostatic or electromagnetic forces offer the potential for dynamic selection capabilities (*22, 36*). However, they face challenges such as the requirement for high voltages in electrostatic transfers and the requirements for materials responsive to magnetic fields in electromagnetic transfers, which limit their widespread use and scalability (*37*). Fluidic self-assembly, which uses gravitational and capillary forces for microchip positioning, shows promise in microLED display fabrication due to its rapid and efficient assembly process (*23, 24*). However, this method lacks the dynamic configurability necessary for precise repairs of microLED chips, which is crucial for managing display pixel yields during microLED display manufacturing (*38, 39*). Additionally, the requirement for shape matching between chips and substrate cavities restricts the diversity of devices that can be assembled, posing further application challenges. Laser-driven release provides a solution by selectively releasing chips through a laser beam that induces thermal effects like deformation, adhesion modification, or material phase changes at the interface between the dies and the carrier (*6, 19, 20, 40*). While this technology enables precise device targeting and dynamic component selection, it risks material damage from high temperatures and involves high costs due to complex laser systems, limiting broader adoption. Additionally, the requirement for accurate gap control between stamps and receiver substrates poses a significant challenge for non-contact release methods (*37*).

Systems using shape memory polymers (SMPs) offer a promising solution with their ability to handle arbitrarily shaped objects and moderate actuation conditions (*41-43*). These systems use thermally activated SMP blocks to modulate adhesion through changes in modulus and shape recovery, enabling selective picking and releasing of micro-object arrays via localized heating (*41, 44-48*). Laser-induced heating has been explored in early studies as a method to locally activate SMPs for selective release of microchiplets (*48*). While such approaches continue to be employed in recent publications highlighting selective transfer functionalities (*41, 44, 45*), they are fundamentally limited by the inherently sequential nature of laser scanning, resulting in low transfer throughput for large-area applications. The challenges such as gap-induced registration errors and inability to handle nano-thick materials also limit their broader applications (*45*). Moreover, laser-driven approaches also depend on high-end optical and laser systems, which can substantially raise system complexity and cost, thereby affecting manufacturing cost. A prior study demonstrated a basic proof of concept for using microheaters to locally activate SMPs for selective transfer of silicon dies (*46*). However, the work remained at an early conceptual stage, limited to a 3×3 array of 50 μm silicon dummy chips, without addressing several critical aspects necessary for practical application. In particular, the study did not address key issues such as the diversity of transferable materials, compatibility with functional device interfaces, and integration scalability—factors that are essential for enabling reliable heterogeneous and multi-material assembly in in-plane or 3D integration.

In this work, we present a fully developed and versatile μTP platform that achieves significant advances in material functionality, transfer diversity, and integration capability. Through systematic material optimization and device/system design, we establish a broadly applicable transfer printing framework capable of reliably assembling semiconductors, polymers, metals, nanometer-thick films, and microspheres—demonstrating integration capabilities that have not been achieved by previous transfer approaches. Importantly, we demonstrate unprecedented integration capabilities, including programmable 3D stacking and suspended vertical assembly of nanometer-thick materials—laying the groundwork for constructing next-generation microsystems. The system effectively handles exceptionally small components, including microchiplets as small as 23 μm × 23 μm and 2 μm thick, as well as thin films only 10 μm wide and 90 nm thick. It also supports a broad range of material forms—from microchiplets to nanometer-scale thin films and microspheres—offering flexible manufacturing capabilities for diverse applications. We have also successfully achieved the dynamically programmable transfer of 45 μm × 25 μm, 8-μm-thick microLED chips to create arbitrary patterns with high precision, demonstrating the technology's potential for mass transfer and smart repair processes, addressing a critical bottleneck in the manufacturability and commercialization of microLED technology. These results suggest the significant potential impact of this technology in revolutionizing manufacturing processes for advanced electronic and photonic devices.

## Results

### ***Principle of the dynamically programmable micro-transfer printing system***

Figure 1A shows the schematic of the transfer head in the dynamically programmable μTP technology. It consists of the sharp phase-changing rigid-to-rubbery (SPRR) polymer for adhesion modulation, an independently addressable microheater array, and a glass substrate. The SPRR polymer, primarily composed of stearyl acrylate (SA) and long-chain urethane diacrylate (UDA), exhibits a sharp change in storage modulus from approximately 130 kPa to about 230 MPa across its melting temperature ($T_m$) of 44 °C (see Supplementary fig. S1). This change represents a nearly 1800-fold change in modulus containing a sharp transition over a narrow temperature range of less than 10 °C. The microheater array consists of 30-nm thick ITO resistors with 5 μm wide, curved traces that form microheaters with approximately circular outlines, each 40 μm in diameter. Electrical connections are made using Ti/Au interconnects with thicknesses of 100 nm and 300 nm, respectively. The array has a pitch of 100 μm. The microheaters can be individually controlled to locally adjust the adhesion of the SPRR polymer, enabling dynamically programmable transfer printing.

The principle of the dynamically programmable transfer process is illustrated in Fig. 1B. The pickup and release are dictated by the fracture competition between two interfaces: the SPRR polymer/micro-object (P/O) and the micro-object/substrate (O/S) interfaces. Separation occurs at the weaker interface with lower energy release rate, when the substrate is moved from the transfer head. During the pickup process, the transfer head first contacts the micro-objects on the donor substrate and heats specific areas of the SPRR polymer to a rubbery state using microheaters. The SPRR polymer heated then achieves complete conformal contact with the selected micro-objects' surfaces. Upon cooling to room temperature, the SPRR polymer transitions to a rigid state, maintaining this conformal contact and exhibiting strong adhesion to the selected micro-objects due to interface locking. At this point, the interface between the selected micro-objects and the rigid SPRR polymer has maximized energy release rate $G_{crit}^{(P/O)}$ due to the high-module interface system (*49, 50*), which is greater than that of the micro-object/substrate interface ( $G_{crit}^{(O/S)}$ ). Consequently, the transfer head picks up these selected micro-objects upon separation from the donor substrate. In the meantime, regions of the SPRR polymer that have not undergone the heating-contact-cooling process remain rigid, thus preventing the pickup of unselected micro-objects due to inadequate adhesion. During the release process, the SPRR polymer is heated to a rubbery state as the transfer head separates from the receiver substrate. The energy release rate $G_{crit}^{(P/O)}(v)$ of the interface between rubbery SPRR polymer and micro-object is speed-dependent and can be minimized at an appropriate separation speed, facilitating the transfer of micro-objects onto the receiver substrate. This mechanism demonstrated an adhesion strength ratio of pickup force to release force higher than 190:1. Further details on the transfer mechanism and adhesion characterization can be found in Supplementary text and Supplementary fig. S2.

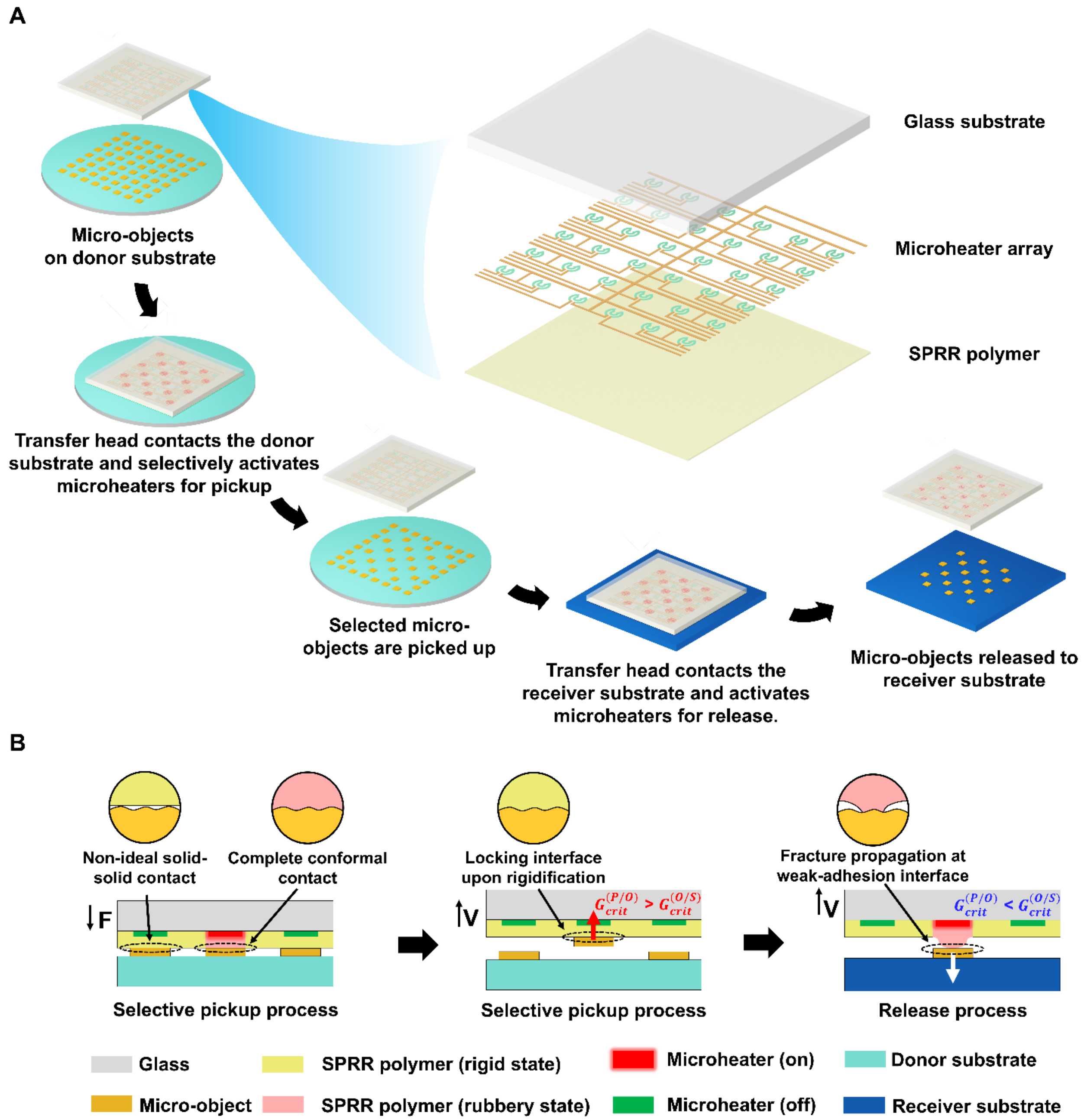

**Fig. 1. The dynamically programmable µTP technology.** (**A**) Schematic of the transfer head in the µTP technology, along with the dynamically programmable transfer process flow. (**B**) Illustration of the dynamically programmable transfer principle, showing the pickup and release protocols.

## ***Transfer head characterization***

The transfer heads were fabricated on glass wafers using microfabrication techniques. The ITO microheater array and metallic interconnects were made through deposition, etching and photolithography-based patterning. A 30-µm-thick SPRR polymer layer was applied using a gap-filling molding process and is solidified by UV curing. The assembly was finalized by wire bonding the metallic interconnections to external cables. Fig. 2A shows a fabricated transfer head assembly mounted on an aluminum fixture. To evaluate the basic functions and characteristics of the transfer head, it was observed under a microscope whiles microheaters were switched on and off. A camera, operating at a 50 Hz sampling rate, captured changes in the SPRR polymer of the actuated pixels as they surpassed the melting temperature ($T_m$) during actuation. A relay multiplexer facilitated control of the microheater array, enabling simultaneous activation of selected microheaters using a single power supply. The power delivered to each microheater can be optimized using external circuits. Fig. 2B displays the transfer head with selected actuated pixels when ten microheaters are simultaneously activated at a total power of 52.7 mW for 150 ms, with individual microheater power ranging from 3.4 mW to 7.7 mW. Clear boundaries around each actuated µTP pixel are observed due to the sharp transition from the rigid to the rubbery state and the transparency differences between the two states of the SPRR polymer. This sharp contrast facilitates precise device targeting during the selective transfer of closely compacted devices, affecting only the intended components while leaving adjacent ones unaltered. Supplementary Movie S1 shows the continuous actuation of selected pixels as they are turned on and off. More details on the fabrication process can be found in Methods and Supplementary fig. S3.

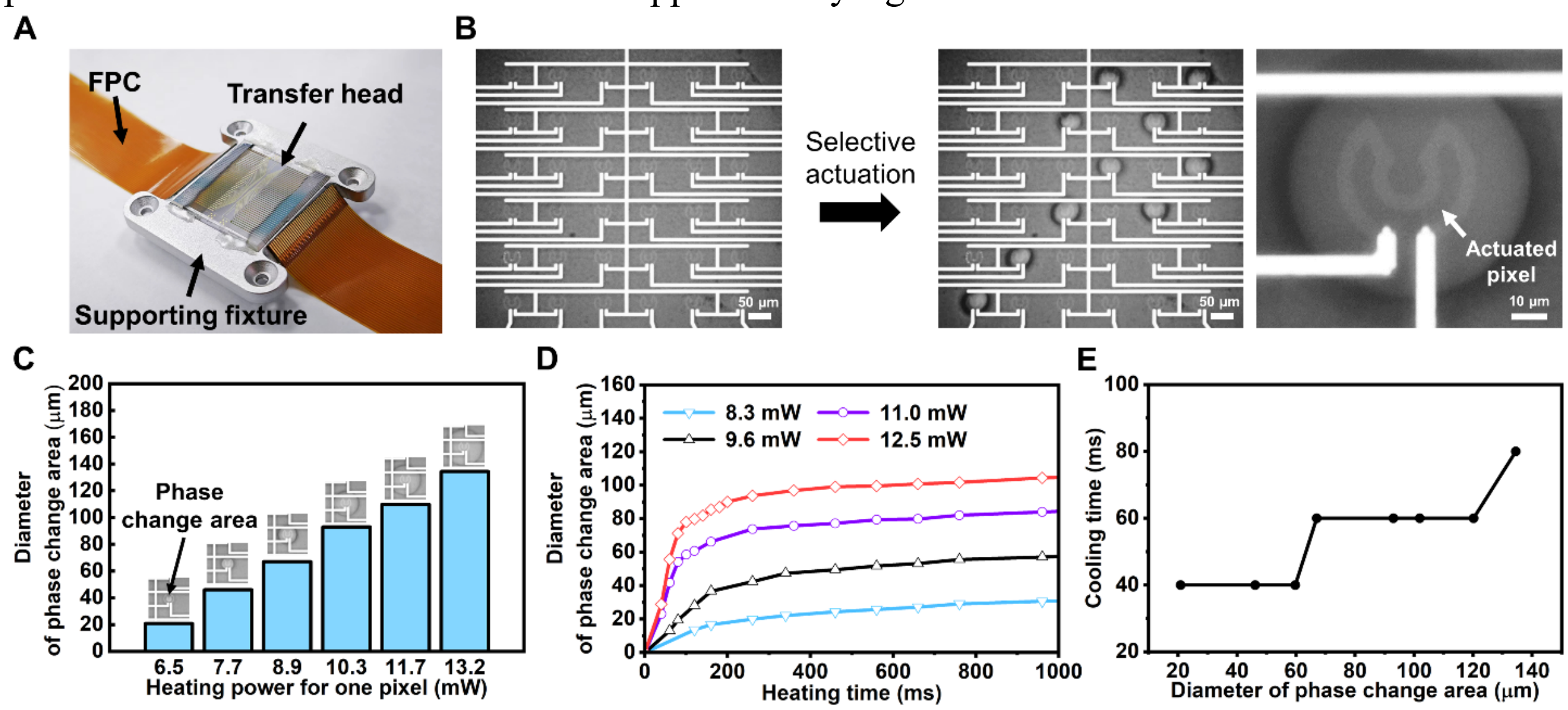

**Fig. 2. Characterization of the transfer head.** (**A**) Photo of the transfer head mounted on a supporting fixture. (**B**) Microscopic images showing a group of ten actuated µTP pixels under a specific heating power. (**C**) The equilibrium phase transition area of an actuated µTP pixel in response to varied heating powers. (**D**) The change of µTP pixel's phase transition area as a function of heating time when µTP pixel is actuated under different heating powers. (**E**) The cooling times required for a heated µTP pixel to return to its rigid state varying depending on its phase transition area. Note the time measurements in (D) and (E) have an uncertainty of ±20 ms due to the limited capture sampling rate, which is 50 Hz.

To characterize the thermal response of the µTP pixel, a single pixel was actuated with varying levels of heating power. During actuation, an LED indicator with nanosecond-scale latency was connected in parallel with the microheater. The illumination changes of this LED in the microscope display served as a timestamp, indicating the precise moments when heating began and ended. The phase change process of the µTP pixel was recorded using the microscope camera at a 50 Hz sampling rate. Fig. 2C illustrates the equilibrium phase change area of the 30-µm-thick SPRR polymer in the transfer head as the actuated pixel responds to varying heating powers. This demonstrates the correlation between actuation power and the equilibrium size of the polymer's phase transition area. Dynamically adjusting the heating power allows for precise control of the phase transition area for µTP. The heating and cooling times required for phase transitions in a specific area determine the potential drive speed of each pixel in transfer printing processes. These times, along with the phase change area of the μTP pixel, were determined by analyzing LED illumination duration and examining the phase change area recorded by the video. Fig. 2D illustrates how the phase transition area in a 30-μm-thick SPRR polymer varies with heating time under different heating powers. The result indicates that increasing the heating power enables faster actuation. A phase transition occurs across a 50-μm-diameter region within just 60±20 ms at a power of 12.5 mW. Fig. 2E shows the cooling times needed for a heated μTP pixel to return to a rigid state. Specifically, a heated pixel with a phase transition diameter of approximately 60 μm returns to a rigid state in 40±20 ms through natural cooling under ambient conditions. The uncertainty in these time measurements is due to the limited capture sampling rate which is 50 Hz. These results indicate the potential for rapid pixel actuation during μTP process. In the study, multiple transfer heads with different microheater configurations were fabricated. Further details on these configurations can be found in Supplementary fig. S4.

***Versatile material assembly and heterogeneous integration***

To evaluate system performance and demonstrate dynamically programmable transfer, we conducted experiments with various micro- and nano-scale materials and components. These objects were transferred using a setup equipped with a six-axis stage (Supplementary fig. S5), and the process was carefully monitored under an optical microscope while being controlled through precise heating/cooling protocols and stage motion. Fig. 3A and Supplementary Movie S2 illustrate the dynamically programmable transfer process and results of 50-μm-by-50-μm, 3-μm-thick AZ5214E photoresist (PR) chiplets. First, the chiplet array with a pitch of 100 μm was prepared by doing photolithography on the donor substrate after surface hydrophilic treatment. A subset of this array was then selectively picked up by the transfer head and released onto a receiver substrate. During the pickup phase of the experiment, the selected group of ten microheaters was simultaneously activated with a total heating power of 80.1 mW for 150 ms, followed by natural cooling. The power applied to each microheater was optimized, ranging from 5.1 mW to 11.7 mW, and the substrate separation speed was set at 5 μm/s. During the release phase, the overall microheater array was simultaneously activated with a total heating power of 261.3 mW and the substrate separation speed was set at 5 μm/s. Fig. 3B shows an overlaid processed microscope image of the microchiplets before and after the transfer where the registration with linear transform was achieved with the assistance of Fiji software (*51, 52*). The registration error, indicating the positional differences between the overlaid images of the chiplets before and after transfer, was assessed. The local translation of the microchiplets' geometric center is less than 0.7±0.5 μm, and local rotation is under 0.04±0.01 rad across an area of 550 μm × 550 μm. Fig. 3C further demonstrates the assembly capability of dynamically programmable transfer, showing a "HKUST (GZ)" pattern formed using AZ5214E photoresist chiplets, transferred from multiple donor arrays. Note that in the described transfer experiments, both the donor and receiver substrates have the same surface material (6.5 mil PF-X4 gel-film, Gel-Pak), demonstrating that this technique can effectively repopulate devices on an identical substrate. Further details about the preparation of AZ5214E chiplets, experimental results and registration error analysis are provided in the Methods, Supplementary fig. S6, and Supplementary text, respectively.

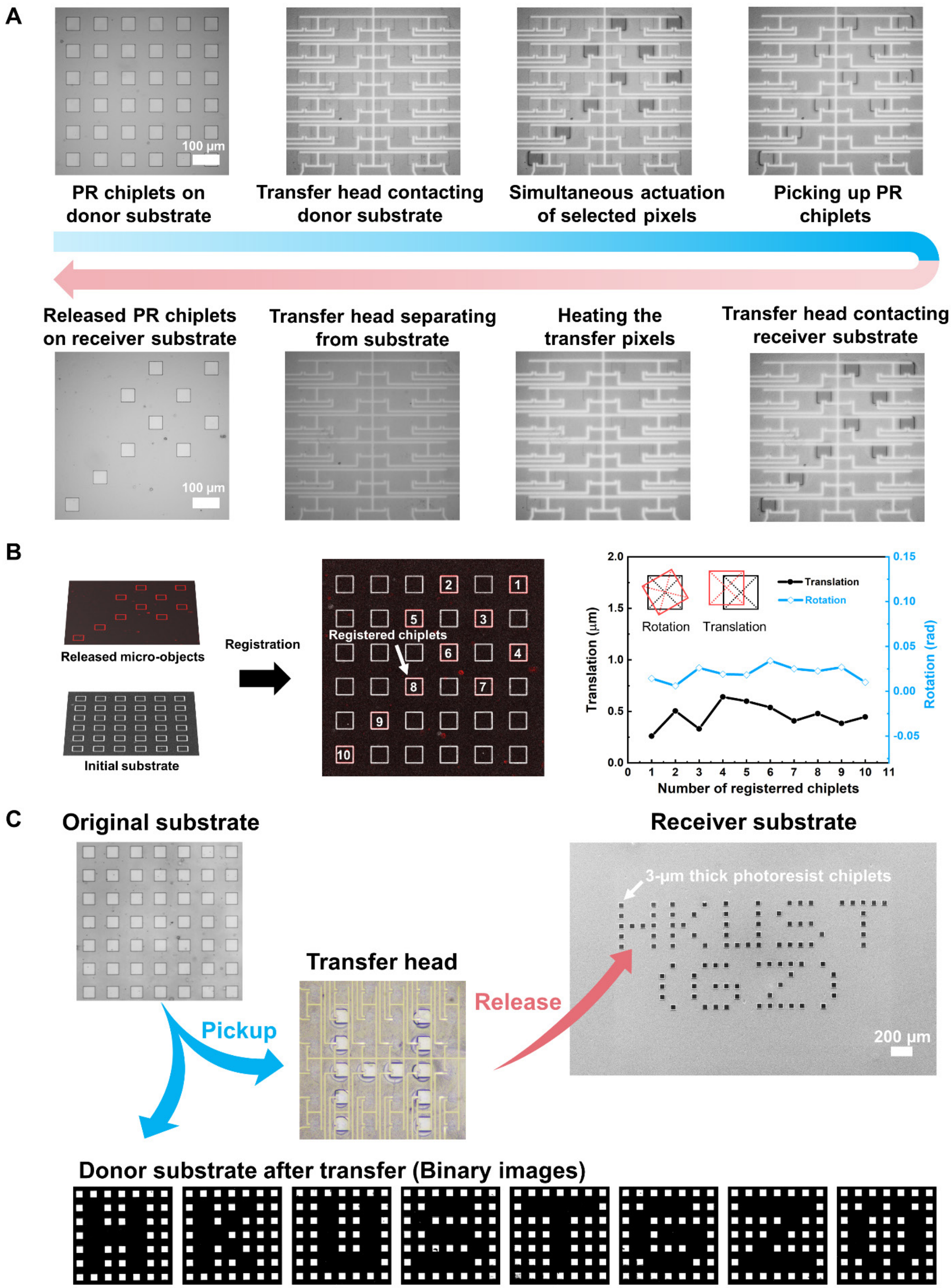

**Fig. 3. Dynamically programmable micro-transfer printing of microchiplets.** (**A**) Transfer process of the selected 3-µm-thick PR microchiplets moving from a donor substrate (6.5 mil PF-X4 gel-film, Gel-Pak) to a receiver substrate (6.5 mil PF-X4 gel-film, Gel-Pak). (**B**) Registration error of microscope images of PR microchiplets before and after transfer. (**C**) PR microchiplets forming the “HKUST (GZ)” pattern using the dynamically programmable µTP process.

In this study, we also demonstrated the applicability of this technology for a wide variety of materials, including photoresists, polyimide, silicon, copper, solder (Sn96.5Ag3.0Cu0.5), polystyrene (PS), and GaN microLED chips. These materials span diverse geometries and dimensions such as micro-scale chiplets, nanometer-scale thin films, and micro-spheres. Further details about the multiple selective transfer demonstrations are provided in Supplementary fig. S7, Supplementary fig. S8 and Supplementary Movie S3. Fig. 4A illustrates the result of selective transfer of 90 nm thick, 50 μm diameter copper films onto a 6.5 mil PF-X4 gel-film and Supplementary fig. S8 shows the donor substrate before and after transfer. Fig. 4B demonstrates the selective transfer result of silicon chiplets with higher aspect-ratio (area: 50 μm × 50 μm, thickness: 70 μm). Leveraging the SMP characteristics of the SPRR polymer, the transfer head effectively handles non-flat objects like micro-spheres, with results for 50-μm-diameter PS micro-spheres and 100-μm-diameter solder ball shown in Fig. 4C and Supplementary fig. S7, respectively. These results indicate that the technology has potential for diverse manufacturing applications where precise manipulation of various geometries is essential.

Furthermore, we conducted experiments to explore the technology's capabilities in stacking and heterogeneously integrating various materials. Fig. 4D shows an example of in-plane assembly, demonstrating tightly packed arrangements of various micro-objects with different thicknesses, including copper nano-film (90 nm thick), AZ5214E photoresist structure (3 μm thick), polyimide (13 μm thick), SU-8 3025 photoresist (18 μm thick), and PS microsphere (50 μm in diameter). Fig. 4, E and F demonstrate out-of-plane stacking experiments of discrete AZ5214E photoresist structures on a supporting “platform” made of AZ5214E photoresist. Fig. 4E shows the precise assembly result of 25 μm × 25 μm AZ5214E photoresist chiplets, each 3 μm thick, stacked with each layer rotated 10°. The contact area between two adjacent photoresist chiplets is smaller than (approximately 0.77 times) the area of a single chiplet, facilitated by the high pickup and release strength ratio of the transfer head. Similarly, Fig. 4F shows a multi-layer structure created by stacking AZ5214E PR layers. Two rectangular thin plate layers (each 48 μm × 8 μm, 2 μm thick) are placed on top of two supporting “piers”, each composed of nine thin plates (each 23 μm × 23 μm, 2 μm thick). Fig. 4G presents an experimental result where two cuboid AZ5214E PR bars (5 μm × 30 μm, 3 μm thick) are first transferred onto the target substrate (6.5 mil PF-X4 gel-film, Gel-Pak) to serve as supporting “piers”. Subsequently, a copper nano-film (10 μm × 100 μm, 90 nm thick) is transferred to span across the top of two separate PR bars, creating a “bridge” structure with nanometer thickness. Fig. 4H shows a 90-nm-thick copper nano-film transferred onto the surface of a solder ball with a diameter of 100 μm. Further details on different donor substrate preparation are in the Methods. The related transfer experiments are also shown in Supplementary Movie S4.

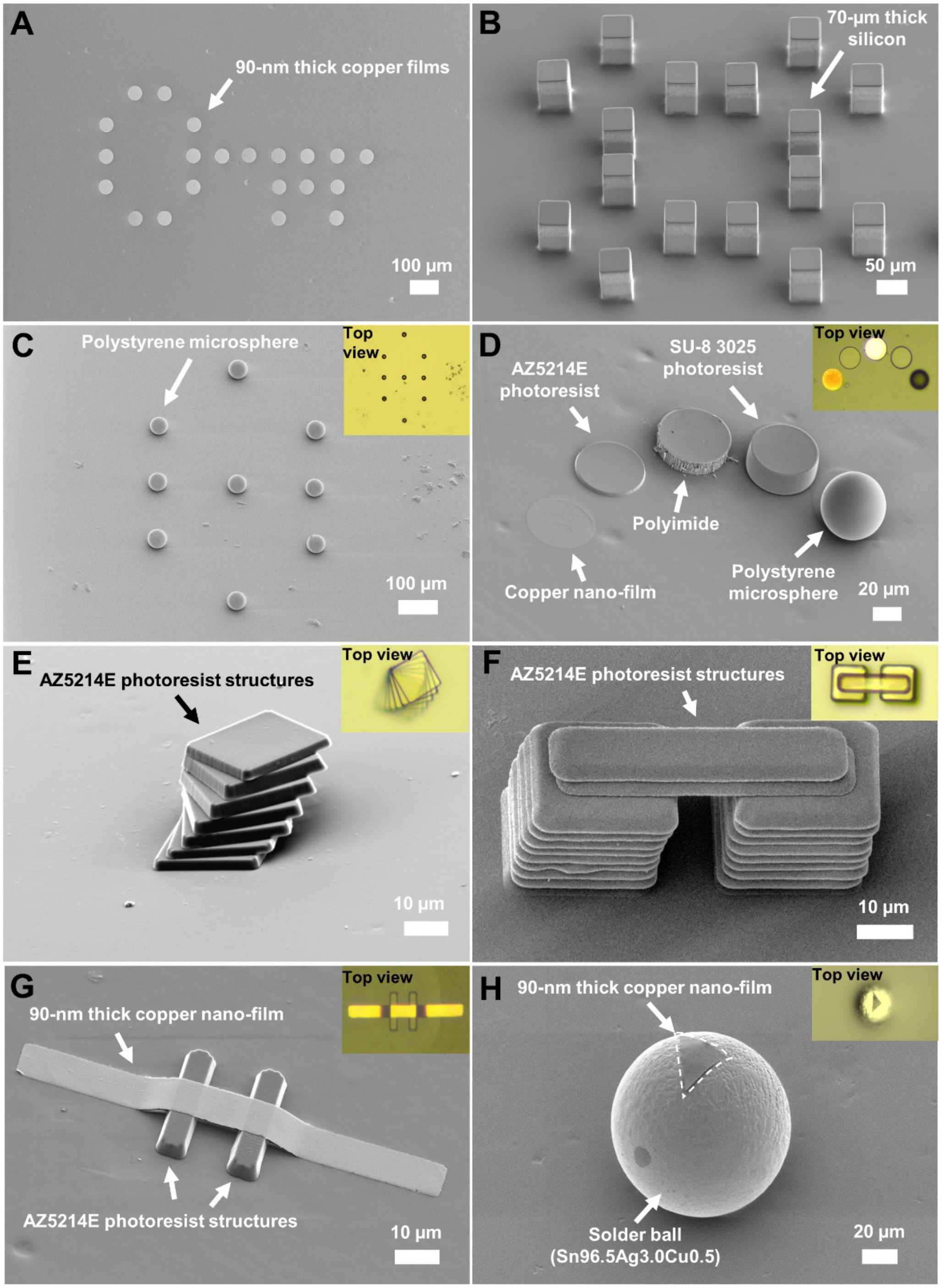

**Fig. 4. Versatile material compatibility and assembly.** (**A**) Selectively transferred 90-nm-thick, 50-µm-diameter copper films on a receiver substrate (6.5 mil PF-X4 gel-film, Gel-Pak). **(B)** Selectively transferred 70-µm-thick, 50 µm × 50 µm silicon chiplets. **(C)** Transfer result of 50-µm-diameter PS microspheres. **(D)** In-plane assembly of multiple heterogeneous materials on the silicone gel-film (6.5 mil PF-X4 gel-film, Gel-Pak), including 90-nm-thick copper nano-film, 3-µm-thick PR (AZ5214 E) structure, 13-µm-thick polyimide, 18-µm-thick SU-8 3025 photoresist, and 50-µm diameter polystyrene microsphere. **(E)** Stacking of 3-µm-thick AZ5214E PR chiplets with each layer rotated an angle of 10 degrees relative to the adjacent layer. (**F**) Out-of-plane stacking of multiple PR (AZ5214E) layers: Two PR thin plates placed on two "piers" stacked with nine PR (AZ5214E) thin plates. **(G)** 90-nm-thick copper nano-film transferred on the top of two separate 3-µm height PR (AZ5214E) bars. (**H**) 90-nm-thick copper nano-film transferred onto a 100-µm-diameter solder ball (Sn96.5Ag3.0Cu0.5).

## ***Dynamically programmable transfer of microLED chips***

MicroLED display offers exceptional brightness, energy efficiency, and high resolution, making it highly promising for a wide range of applications (*8, 23, 24, 53-56*). However, its commercial viability is hindered by significant manufacturing challenges, particularly in the efficient mass transfer and repair of the tiny LEDs (*37, 39, 53*). Defective pixels having arbitrary patterns present huge challenges to efficient sorting and repairing during microLED display manufacturing. Overcoming these technical obstacles is essential for the widespread adoption of microLED displays. In this study, we applied the µTP technology to microLED chips sourced from a leading industry vendor to demonstrate its potential for efficiently assembling and repairing microLED arrays. In one experiment, the microLED chips, each has a size of 45 µm × 25 µm and a thickness of 8 µm, were first released through laser-driven release method from their original sapphire substrate and arranged with a 100 µm pitch on a donor substrate that had a silicone surface (6.5 mil PF-X4 gel-film, Gel-Pak). As shown in Fig. 5A and Supplementary Movie S5, we then conducted a dynamically programmable transfer experiment, transferring a selected subset of microLED chips from the donor substrate to a receiver substrate (both substrates are 6.5 mil PF-X4 gel-films, Gel-Pak) to create an “arrow” pattern. The registration result shows that the local translation of the transferred microLED chips is less than 1±0.3 µm and the local rotation is less than 0.01±0.009 rad over a 550 µm × 550 µm area. This indicates a high-precision microLED chip transfer process for creating arbitrary patterns. To verify the functionality of the microLED chips post-transfer, the chips were then transferred to a polyimide substrate, where they were electrically interconnected and subsequently illuminated, as depicted in Fig. 5B. Details of these transfer experiments, transfer accuracy analysis and the interconnection process are documented in Supplementary fig. S7, Supplementary text, Supplementary fig. S9 and Supplementary fig. S10. The results indicate that our dynamically programmable µTP technology could potentially be used to facilitate the management of pixel yield in microLED production through parallel, dynamic replacement of defective pixels.

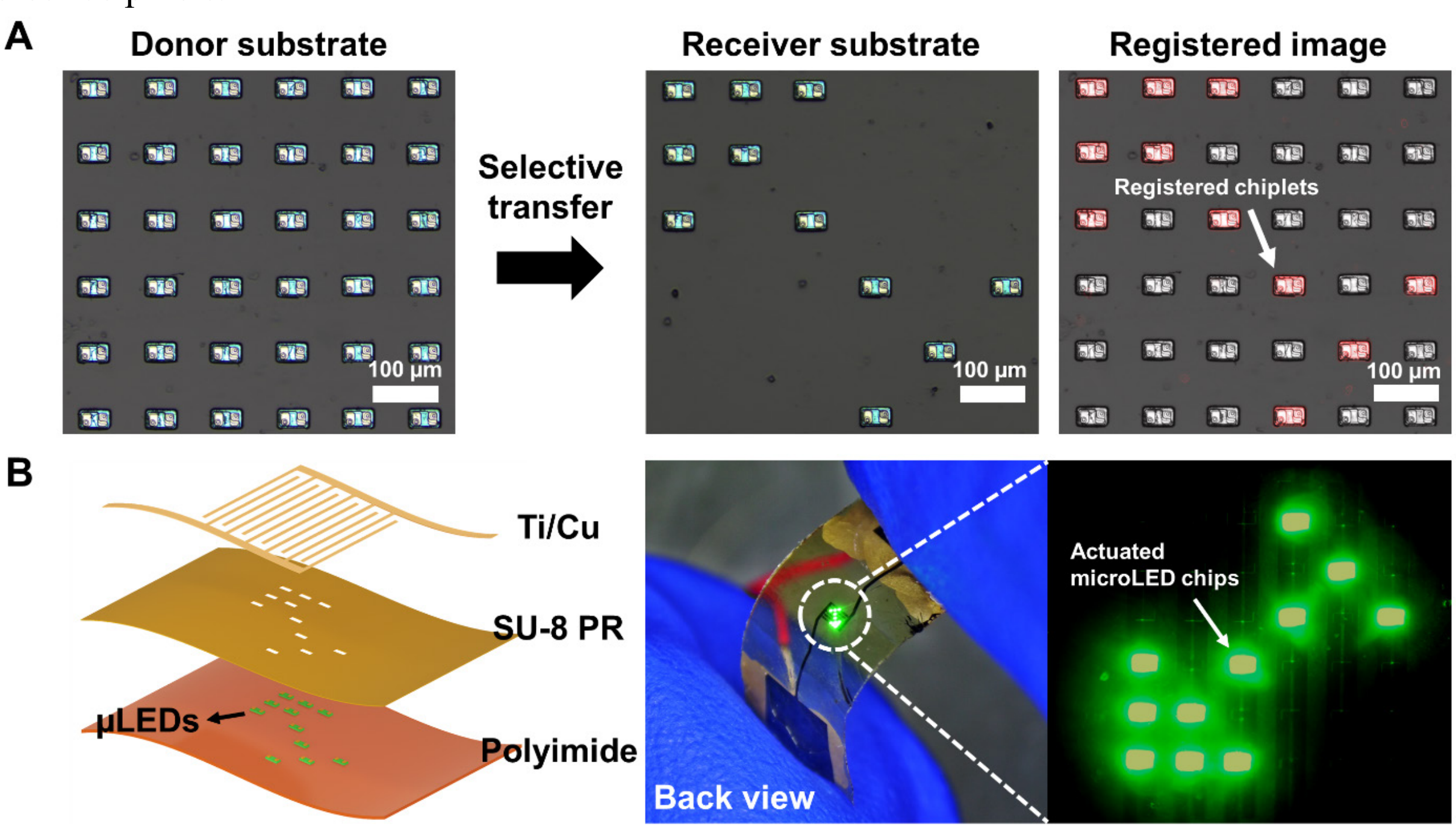

**Fig. 5. Demonstration of the dynamically programmable micro-transfer printing technology for microLED applications** (**A**) Selected microLED chips transferred with an arrow pattern from a donor substrate (6.5 mil PF-X4 gel-film, Gel-Pak) onto a receiver substrate (6.5 mil PF-X4 gel-film, Gel-Pak). (**B**) Schematic of the layered structure (left) and illumination of the flexible microLED signage (right).

## Discussion

We have reported a dynamically programmable micro-transfer printing system that employs a sharp phase-changing rigid-to-rubbery polymer and an independently addressable microheater array. Sharp storage modulus transition characteristics of SPRR polymer and precisely localized heating control by microheaters facilitates dynamically adjustable micro-transfer printing pixels with high resolution for device targeting. Moderate actuation conditions and rapid actuation potential of the micro-transfer printing pixels enable flexible, efficient manufacturing capabilities.

Compared with traditional μTP schemes, our μTP system demonstrates unprecedented dynamically programmable capabilities for selective transfer of various materials, including semiconductors, polymers and metals, handling geometries from micro-scale chiplets to nanometer-thick films and micro-spheres. Our μTP system also demonstrates flexible 3D stacking and heterogeneous integration capability for complex microsystem creation. The general applicability and high transfer accuracy of this dynamically programmable μTP technology unlocks advanced manufacturing and packaging of electronic and photonic systems.

As an application demonstration, we also perform dynamically programmable μTP of microLED chips to create arbitrary pattern with high transfer accuracy. This μTP technology provides a solution to the challenges of pixel yield management in microLED production through parallel, dynamic sorting and replacing defective microLED chips.

## Materials and Methods

### ***Materials***

Urethane diacrylate (UDA) was obtained from Sartomer Company, America. Stearyl acrylate (SA), trimethylolpropane triacrylate (TMP-TA), 2,2-dimethoxy-2-phenyl-acetophenone (DMPA), and benzophenone (BP) were purchased from Sigma-Aldrich, America. Adhesion promoter (3M 94 Primer) was acquired from 3M, America. AR 300-80 new was purchased from Allresist, Germany. AZ5214E photoresist was obtained from AZ Electronic Materials, America. SU-8 2002, SU-8 2010 and SU-8 3025 were acquired from MicroChem, America. Silicone gel-films (6.5 mil PF-X4 gel-film) were purchased from Gel-Pak, America. Trichloro (1H,1H,2H,2H-perfluorooctyl) silane was obtained from Macklin Biochemical Technology Co., Ltd, China. PDMS (SYLGARD 184) was purchased from Dow Corning, America. Polyimide (SMW-610) was purchased from Changzhou Runchuan Plastic Material Co., Ltd, China. 50-μm-diameter polystyrene microspheres were obtained from Suzhou Knowledge & Benefit Sphere Tech. Co., Ltd, China. 100-μm-diameter Sn96.5Ag3.0Cu0.5 solder balls (SAC 305) were acquired from Zhenjiang Fanyada Electronic Technology Co., Ltd, China.

### ***Transfer head fabrication***

The fabrication process of the microheater array is schematically illustrated in fig. S3. First, an adhesion promoter, AR 300-80 new, is spin-coated on the cleaned ITO-coated glass wafer (ITO thickness: 30 nm; glass thickness: 700 μm) at 4000 rpm for 30 s, followed by baking at 180 °C for 2 min. Subsequently, a layer of AZ5214E photoresist is spin-coated on the ITO-coated glass wafer at 4000 rpm for 30 s, and baked at 100 °C for 90 s. Following this, the photoresist is exposed to 365-nm UV light (intensity: 15 $mW/cm^2$, duration: 6 s, SUSS MA/BA6 Gen4) and developed to create pattern. After baking the patterned photoresist at 120 °C for 80 s, the ITO is etched in the IBE instrument (SHL FA100S-IBE) with $Ar^+$ ions for 6 min at a 90° incident angle under 300 V beam voltage and 200 mA beam current. Next, the photoresist is striped in acetone solution to obtain ITO resistors. To prepare the metal traces, the patterned ITO-coated glass wafer is spin-coated again with a layer of AZ5214E photoresist at 4000 rpm for 30 s, followed by a soft baking at 100 °C for 90 s and exposure to 365-nm UV light (intensity: 15 $mW/cm^2$, duration: 4 s, SUSS

MA/BA6 Gen4). The exposed photoresist is then baked at 100 °C for 100 s and flood-exposed under 365-nm UV light (intensity: 15 mW/cm$^2$, duration: 45 s, SUSS MA/BA6 Gen4) for the image reversal process of AZ5214E photoresist. After that, the photoresist is patterned in the developer. Then, 100 nm Ti and 300 nm Au metal layers are sputtered successively onto the ITO-coated glass wafer, after which the photoresist is removed in acetone solution to obtain a resistive micro-heater array on the glass wafer.

The SPRR polymer is customized utilizing a bistable electroactive polymer reported in previous study (*57*). First, a liquid SPRR prepolymer is prepared by stirring a mixture of UDA, SA, TMP-TA, DMPA, and BP in a weight ratio of 40:160:2:2:1 at 80 °C. Before applying the SPRR polymer layer, the microheater array is coated with a layer of adhesion promoter (3M 94 Primer) to enhance its surface adhesion. Then, the liquid prepolymer is carefully dispensed between the microheater array and a smooth cover glass plate which are separated by a 30-μm-thick Kapton tape spacer. The prepolymer is then cured using 300 mW/cm$^2$ UV light for 3 min. After UV curing, the sample is heated on a hot plate at 80 °C for 2 hours to remove residual unreacted compounds. Finally, the transfer head is obtained by removing the cover glass plate at 80 °C.

The fabricated transfer head is then machined with a 45° sidewall using saw dicing process. To facilitate the wire-bonding process, 100 nm Ti and 800 nm Al metal layers are successively sputtered onto the edge region of transfer head, extending from the Ti/Au pad of the microheater array to the sidewall, followed by a laser cutting process to prepare discrete Ti/Al bond pads on the sidewall. To assemble the transfer head into μTP system, both the transfer head and external cable are mounted on an aluminum supporting fixture. Wire-bonding process is then performed to make an electrical connection between external cables (flexible printed circuit) and the transfer head.

***Preparation of micro- to nano-scale components on donor substrates***

*Silicone gel-film/glass substrate:*

First, a commercial glass substrate is sequentially cleaned in acetone, isopropanol and ethanol solution. Following this, a piece of commercial silicone gel-film (6.5 mil PF-X4 gel-film) is laminated on the glass wafer, providing temporary adhesion to the micro-objects.

*AZ5214E photoresist structures:*

First, a silicone gel-film/glass substrate undergoes oxygen plasma treatment (LEBO Science, PT500) at 100 W for 1 min to enhance surface wettability. Following this, a layer of AZ5214E photoresist is spin-coated onto the substrate at 3000 rpm for 30 s and then baked at 100 °C for 100 s, followed by exposure (365-nm UV light, intensity: 15 mW/cm$^2$, duration: 6 s, SUSS MA/BA6 Gen4) and development to create the desired photoresist structures on the silicone gel-film/glass substrate.

*SU-8 2010 photoresist structures:*

First, a silicone gel-film/glass substrate undergoes 1-min oxygen plasma treatment (LEBO Science, PT500) at 100 W. Subsequently, a layer of SU-8 2010 photoresist is spin-coated on the silicone gel-film at 3000 rpm for 30 s, followed by baking at 65 °C for 1 min and then at 95 °C for 3 min. The SU-8 2010 photoresist is then exposed to 365-nm UV light (SUSS MA/BA6 Gen4) at 15 mW/cm$^2$ for 7 s, followed by baking at 65 °C for 1 min and then at 95 °C for 3 min. Finally, the SU-8 2010 photoresist is developed to obtain desired structures on the silicone gel-film/glass substrate.

*SU-8 3025 photoresist structures:*

First, a silicone gel-film/glass substrate undergoes oxygen plasma treatment (Tergeo-Plus, PIE Scientific) at 50 W for 50 s. Subsequently, SU-8 3025 photoresist is spin-coated onto the silicone gel-film at 3000 rpm for 30 s, followed by sequential baking at 65 °C for 1 min and at 100 °C for 5 min. The photoresist is then exposed to 365-nm UV light (URE-2000S/AL) at an intensity of 9 $mW/cm^2$ for 10 s. Following this, the photoresist is baked again at 65 °C for 1 min and then at 100 °C for 5 min. Finally, the photoresist is developed to form patterned SU-8 3025 structures on the silicone gel-film/glass substrate.

*Polyimide structures:*

First, a commercial off-the-shelf free-standing polyimide film (13 µm thick) is laminated onto a silicone gel-film/glass substrate. The polyimide surface then undergoes oxygen plasma treatment (Tergeo-Plus, PIE Scientific) at 50 W for 60 s to enhance surface wettability. After that, AZ5214E photoresist is spin-coated onto the polyimide surface and then patterned. The surface is then sequentially sputtered with 30-nm-thick Cr and 100-nm-thick Al, followed by photoresist removal using an acetone solution. Finally, the polyimide is etched in a reactive ion etching (SHL FA210E-RIE) chamber at a RF bias power of 200 W for 20 min with $O_2$ /$Ar_2$ gas mixture (flow rate: 250/50 sccm, pressure: 278.3 mTorr) to obtain discrete polyimide structures on the silicone gel-film/glass substrate.

*Silicon chiplets:*

First, a silicon substrate (70 µm thick) is adhered to a silicone gel-film/glass substrate. A layer of AZ5214E photoresist is then spin-coated onto the silicon substrate and patterned. Finally, the silicon is etched using the deep reactive ion etching (DRIE) process producing discrete silicon chiplets.

*Copper nano-film:*

First, a silicone gel-film/glass substrate undergoes oxygen plasma treatment (Tergeo-Plus, PIE Scientific) at 50 W for 50 s. Subsequently, AZ5214E photoresist is spin-coated and patterned on the substrate. Following this, the substrate with patterned photoresist is coated with a monolayer of Trichloro (1H,1H,2H,2H-perfluorooctyl) silane via thermal evaporation at 80 °C for 30 min. Finally, a 90-nm-thick copper film is deposited on the substrate using electron beam evaporation, after which the photoresist is striped using an acetone solution.

*Polystyrene microspheres*:

First, a PDMS mixture consisting of a base solution and curing agent is prepared at a weight ratio of 30:1. The PDMS prepolymer is poured into a cuboid mold and cured at 80 °C for 1 hour. The PDMS film is then peeled from the mold and laminated onto a cleaned glass substrate. The 50-µm-diameter polystyrene microspheres are then arranged into arrays on the PDMS/glass substrate by using a transfer head to individually pick and place each microsphere.

*Solder balls:*

100-µm-diameter Sn96.5Ag3.0Cu0.5 solder balls are arranged into arrays on the polyimide tape/glass substrate by using a transfer head to individually pick and place each solder ball.

*MicroLED chips:*

The commercial GaN microLED chips, with dimensions of 45 µm × 25 µm × 8 µm, were first released from their original sapphire substrate onto a silicone gel-film/glass substrate one by one by using 248 nm laser pulse with a spot size of 50 µm × 50 µm to ablate the material on the interface between microLED chips and sapphire substrate. This process was achieved with a single 20-nanosecond pulse delivering 300 mJ of energy.

***Fabrication of flexible microLED signage***

The fabrication process begins with the microLED chips that have been transferred onto a silicone gel-film/glass substrate by dynamically programmable micro-transfer printing technology.

Subsequently, a 1 mm thick SPRR polymer block stamps silicone gel-film/glass substrate in rigid state and then both of which are heated to 50 °C on a hot plate. After transitioning into rubbery state, the SPRR polymer block is naturally cooled down to room temperature under ambient conditions. Subsequently, the microLED chips are all picked up by the SPRR polymer block by separating the silicone gel-film/glass substrate from the SPRR polymer block. In order to eliminate the embedding depth of microLEDs within the SPRR polymer block, the SPRR polymer block with transferred microLED chips is heated up to 50 °C again and then cooled down to room temperature. Then, a layer of polyimide prepolymer is spin-coated on the microLED chips at 1000 rpm for 30 s followed by baking at 150 °C for 2 hours. After curing, the polyimide film with microLED chips is separated from the SPRR polymer block in rubbery state and then is laminated on a silicone gel-film/glass substrate in such a way that the electrodes of microLED chips face away from the silicone gel-film/glass substrate. After that, the polyimide film undergoes an oxygen plasma treatment (Tergeo-Plus, PIE Scientific) at 100 W for 120 s to enhance its surface wettability. An adhesion promoter (AR 300-80 new) is then spin-coated on the polyimide film at 4000 rpm for 30 s, and baked at 180 °C for 2 min. Following this, a layer of SU-8 2002 photoresist is spin-coated on the polyimide film at 2000 rpm for 30 s, followed by a baking at 80 °C for 2 min and then at 110 °C for 5 min. The SU-8 2002 photoresist is then exposed to 365-nm UV light (URE-2000S/AL) at 9 mW/cm$^2$ for 11 s, with subsequent baking at 80 °C for 2 min and then at 110 °C for 5 min. The SU-8 2002 photoresist is developed for exposing the electrodes of microLED chips and then hard-baked at 180 °C for 2 hours. A layer of AZ5214E photoresist is then spin-coated on the polyimide film and patterned. Finally, a 50 nm Ti and 120 nm Cu are sputtered onto the polyimide film followed by removal of the AZ5214E photoresist in acetone solution to finish electric interconnection of microLED chips.

**Declaration of generative AI and AI-assisted technologies in the writing process**
During the preparation of this work the authors used ChatGPT in order to refine the English in the manuscript. After using this tool, the authors reviewed and edited the content as needed and take full responsibility for the content of the publication.

**Acknowledgments**

The authors want to thank Prof. Qingming Chen at Sun Yat-sen University and Prof. Shengdong Zhang at Peking University for their valuable discussion. Also, the authors want to acknowledge the support in equipment for fabrication and characterization from the following lab in The Hong Kong University of Science and Technology (Guangzhou): Wave Functional Metamaterial Research Facility (WFMRF), Nanosystem Fabrication Facility (NFF), Materials Characterization and Preparation Facility (MCPF), Advanced Additive Manufacturing Laboratory (AAM), The Center for Heterogeneous Integration of μ-systems and Packaging (CHIP), and Laboratory for Brilliant Energy Science and Technology (BEST Lab). The authors also want to thank the equipment support from the laboratory of School of Microelectronics Science and Technology in Sun Yat-sen University, and Laser Micro/Nano Processing Lab in Guangdong University of Technology. ChatGPT was used to refine the English in the manuscript.

**Funding:**

The National Natural Science Foundation of China (No. 52375580)

The Guangdong Basic and Applied Basic Research Foundation (No. 2024A1515011397)

The Department of Education of Guangdong Province (No. 2023ZDZX1036)

The Guangzhou-HKUST(GZ) Joint Funding Program (No. 2023A03J0688)

The 2024 Guangzhou Basic and Applied Basic Research Scheme (No. 2024A04J6466)

This work is also partially supported by Shenzhen Sitan Technology Co., Ltd.

**Author contributions:**

Conceptualization: YW, QG

Methodology: YW, QG, LY

Investigation: QG, LY, YW, YG, JZ, JZ, JJ, WL, KC, CZ

Data Curation: LY and QG

Visualization: QG and LY

Funding Acquisition: YW

Project administration: YW

Supervision: YW

Writing – original draft: QG, YW

Writing – review & editing: YW, QG, LY, YG, JZ, JZ, JJ, WL, KC, CZ

**Competing interests:** Authors declare that they have no competing interests.

**Data and materials availability:** All data are available in the main text or the supplementary materials.

Supplementary Materials for

# Microassembly of Multi-Material and 3D Integration Enabled by Programmable and Universal High-Precision Micro-Transfer Printing

Qinhua Guo *et al.*

*Corresponding author: Yunda Wang. Email: ydwang@ust.hk

**This PDF file includes:**

Supplementary Text
Figs. S1 to S10

**Other Supplementary Materials for this manuscript include the following:**

Movies S1 to S5

**Supplementary Text**

Transfer mechanism

The switchable adhesion of SPRR polymer fundamentally relies on huge modulus transition, which is driven by the melting-crystallization transition of the SA crystalline aggregates in the SPRR polymer. The SA crystalline aggregates in the SPRR polymer significantly determines the storage modulus change over a narrow temperature range while the long-chain UDA maintains the rubbery phase of SPRR polymer. At room temperature, the SA crystalline aggregates act as hard segments to maintain the high storage modulus of SPRR polymer. When the SPRR polymer temperature rises and exceeds the melting point (~44 °C), SA crystalline aggregates melt and become plasticizers to soften the SPRR polymer, enabling huge modulus reduction over a narrow temperature range.

The transfer mechanism can be explained as follows: the pickup and release are dictated by the fracture competition between the interfaces of SPRR polymer/micro-object and the micro-object/substrate, where the fracture separation occurs in the interface with lower energy release rate. In the pickup processes, the SPRR polymer is first heated over the melting temperature to soften the polymer. The SPRR polymer with low modulus can have conformal contact with the micro-objects. After cooling down to room temperature, the SPRR polymer transitions into rigid state to maintain complete conformal contact with micro-objects. In this case, the SPRR polymer/micro-object interface becomes rigid, remaining a high energy release rate that exceeds the energy release rate of micro-object/substrate interface. When the transfer head retracts from the donor substrate, the micro-objects can be picked up. When the transfer head brings the micro-objects into contact with the receiver substrate, and then heated the SPRR polymer into rubbery state, the SPRR polymer/micro-object interface becomes viscoelastic and has a speed-dependent energy release rate, which can be kinetically controlled below the energy release rate of micro-object/substrate interface. Thus, when the transfer head retracts at an appropriately low separation speed, the micro-objects can be released onto receiver substrate.

To demonstrate the adhesion transition behaviors of SPRR polymer, we performed adhesion measurements and estimations for a SPRR polymer/polyimide interface in different phases using a customized setup equipped with a universal testing machine, a 90° peel test fixture, a 5 N load cell and a hot plate. The hot plate was used to control the temperature of the SPRR polymer and polyimide film. The 90° peel test fixture was linked to the universal testing machine, ensuring synchronization of lateral displacement of test samples with the vertical peeling length when the material is peeled vertically from one end at a constant speed V. The load cell was used to continuously record the peel force $F$ normal to the interfaces. The energy release rate $G$ for the separated interfaces can be calculated by

$$G=\frac{F}{w}$$

where $w$ represents the width of crack propagation. Before performing the peeling tests, 20-mm-width, 13-μm-thick polyimide films were laminated onto a 1-mm-thick SPRR polymer block (with a supportive glass substrate) at 80 ℃. Fig. S2A demonstrates the measurement of kinetically controlled adhesion between the rubbery SPRR polymer and polyimide film at 80 ℃. The energy release rate of the polyimide/SPRR polymer interface rises with peeling speed due to viscoelasticity of rubbery SPRR polymer, showing a small value of ~0.5 $J/m^2$ at a low peeling speed of 0.2 mm/s. Fig. S2B shows tape peeling tests on the polyimide film that was firmly locked by the SPRR polymer at room temperature. It demonstrates a high energy release rate of the tape/polyimide interface, reaching up to ~95 J/m$^2$ at a peeling speed of 8 mm/s. According to fracture competition mechanism, the energy release rate of the polyimide/SPRR polymer interface is higher than that of tape/polyimide interface when the separation occurs at the tape/polyimide interface. Thus, the measurement in Fig. S2B can be utilized to roughly estimate the range of the adhesion strength between polyimide and rigid SPRR polymer. It can be inferred that SPRR polymer has an adhesion switching ratio of pickup to release higher than 190:1.

<u>Transfer accuracy analysis</u>

The images of micro-objects before and after transfer are overlaid with the assistance of Fiji software through linear stack alignment with SIFT (*51, 52*). The linear transform just involves global translation, global rotation and scaling of images while not nonlinearly affecting local information within images. The transfer error here is defined with local translation and local rotation of the transferred micro-objects, which can be measured in the registered images. The translation of each micro-object here is defined as the displacement of their geometric center post-transfer and rotation is defined as the angle by which each micro-object rotates around its geometric center after transfer.

Therefore, the translation vector of each transferred micro-objects can be expressed as:

$$(x, y) = (x_2 - x_1, y_2 - y_1) = (x_2, y_2) - (x_1, y_1)$$

where $(x_1, y_1)$ and $(x_2, y_2)$ represent geometric center coordinates of the micro-object before and after transfer, respectively. Consequently, the registration error from translation can be expressed as:

$$|(x, y)| = \sqrt{(x_2 - x_1)^2 + (y_2 - y_1)^2}$$

In the registered images, the equations for each micro-object are as follows:

$$a_1 \cos \alpha_1 + b_1 \sin \alpha_1 = A_1$$
$$b_1 \cos \alpha_1 + a_1 \sin \alpha_1 = B_1$$
$$a_2 \cos \alpha_2 + b_2 \sin \alpha_2 = A_2$$
$$b_2 \cos \alpha_2 + a_2 \sin \alpha_2 = B_2$$

where $a_1$ and $b_1$ respectively denote the width and length of the micro-object before transfer. $a_2$ and $b_2$ respectively represent the width and length of the micro-object after transfer. $\alpha_1$ and $\alpha_2$ are the angles of the micro-objects before and after transfer, respectively. $A_1$ and $A_2$ respectively denote the width of the minimum exterior rectangles before and after transfer in which the width is along the ***y***-axis. $B_1$ and $B_2$ respectively represent the length of the minimum exterior rectangle before and after transfer and the length is along the ***x***-axis.

Following equations can be further derived:

$$(a_1^2 + b_1^2) \sin \alpha_1 \cos \alpha_1 + a_1 b_1 = A_1 B_1$$
$$(a_2^2 + b_2^2) \sin \alpha_2 \cos \alpha_2 + a_2 b_2 = A_2 B_2$$

The rotation angle of transferred micro-objects can be expressed as follows:

$$
\begin{aligned}
&\alpha(A_1, B_1, A_2, B_2, a_1, b_1, a_2, b_2) \\
&= \alpha_2 - \alpha_1 \\
&= 0.5 \arcsin\left[\frac{2(A_2B_2 - a_2b_2)}{a_2^{\,2} + b_2^{\,2}}\right] - 0.5 \arcsin\left[\frac{2(A_1B_1 - a_1b_1)}{a_1^{\,2} + b_1^{\,2}}\right]
\end{aligned}
$$

Using the Fiji software, the $(x_1, y_1)$, $(x_2, y_2)$, $A_1$, $B_1$, $A_2$, $B_2$, $a_1$, $b_1$, $a_2$, $b_2$ can be measured, and thus the translation error and local rotation angle can be obtained with above equations.

The measurement errors are also analyzed here. In the image registration analysis, the smallest pixel size of images is ε, and thus the uncertainty of distance measurement is $\pm\,\varepsilon$. The maximum estimated error in the translation measurements of micro-objects before and after transfer can be expressed as follows:

$$
\varepsilon_{|(x,y)|} = \pm\sqrt{\left[\sqrt{(x+\varepsilon)^2 + y^2} - \sqrt{x^2 + y^2}\right]^2 + \left[\sqrt{x^2 + (y+\varepsilon)^2} - \sqrt{x^2 + y^2}\right]^2}
$$

The measurement error in the rotation angle can be expressed as follows:

$$
\begin{aligned}
\varepsilon_\alpha^2 = &\left[\alpha(A_1+\varepsilon, B_1, A_2, B_2, a_1, b_1, a_2, b_2) - \alpha(A_1, B_1, A_2, B_2, a_1, b_1, a_2, b_2)\right]^2 \\
&+ \left[\alpha(A_1, B_1+\varepsilon, A_2, B_2, a_1, b_1, a_2, b_2) - \alpha(A_1, B_1, A_2, B_2, a_1, b_1, a_2, b_2)\right]^2 \\
&+ \left[\alpha(A_1, B_1, A_2+\varepsilon, B_2, a_1, b_1, a_2, b_2) - \alpha(A_1, B_1, A_2, B_2, a_1, b_1, a_2, b_2)\right]^2 \\
&+ \left[\alpha(A_1, B_1, A_2, B_2+\varepsilon, a_1, b_1, a_2, b_2) - \alpha(A_1, B_1, A_2, B_2, a_1, b_1, a_2, b_2)\right]^2 \\
&+ \left[\alpha(A_1, B_1, A_2, B_2, a_1+\varepsilon, b_1, a_2, b_2) - \alpha(A_1, B_1, A_2, B_2, a_1, b_1, a_2, b_2)\right]^2 \\
&+ \left[\alpha(A_1, B_1, A_2, B_2, a_1, b_1+\varepsilon, a_2, b_2) - \alpha(A_1, B_1, A_2, B_2, a_1, b_1, a_2, b_2)\right]^2 \\
&+ \left[\alpha(A_1, B_1, A_2, B_2, a_1, b_1, a_2+\varepsilon, b_2) - \alpha(A_1, B_1, A_2, B_2, a_1, b_1, a_2, b_2)\right]^2 \\
&+ \left[\alpha(A_1, B_1, A_2, B_2, a_1, b_1, a_2, b_2+\varepsilon) - \alpha(A_1, B_1, A_2, B_2, a_1, b_1, a_2, b_2)\right]^2
\end{aligned}
$$

$$
\begin{aligned}
\varepsilon_\alpha &\approx \pm\sqrt{\left(\frac{\partial\alpha}{\partial A_1}\right)^2\varepsilon^2 + \left(\frac{\partial\alpha}{\partial B_1}\right)^2\varepsilon^2 + \left(\frac{\partial\alpha}{\partial A_2}\right)^2\varepsilon^2 + \left(\frac{\partial\alpha}{\partial B_2}\right)^2\varepsilon^2 + \left(\frac{\partial\alpha}{\partial a_1}\right)^2\varepsilon^2 + \left(\frac{\partial\alpha}{\partial b_1}\right)^2\varepsilon^2 + \left(\frac{\partial\alpha}{\partial a_2}\right)^2\varepsilon^2 + \left(\frac{\partial\alpha}{\partial b_2}\right)^2\varepsilon^2} \\
&= \pm\varepsilon\sqrt{\left(\frac{\partial\alpha}{\partial A_1}\right)^2 + \left(\frac{\partial\alpha}{\partial B_1}\right)^2 + \left(\frac{\partial\alpha}{\partial A_2}\right)^2 + \left(\frac{\partial\alpha}{\partial B_2}\right)^2 + \left(\frac{\partial\alpha}{\partial a_1}\right)^2 + \left(\frac{\partial\alpha}{\partial b_1}\right)^2 + \left(\frac{\partial\alpha}{\partial a_2}\right)^2 + \left(\frac{\partial\alpha}{\partial b_2}\right)^2}
\end{aligned}
$$

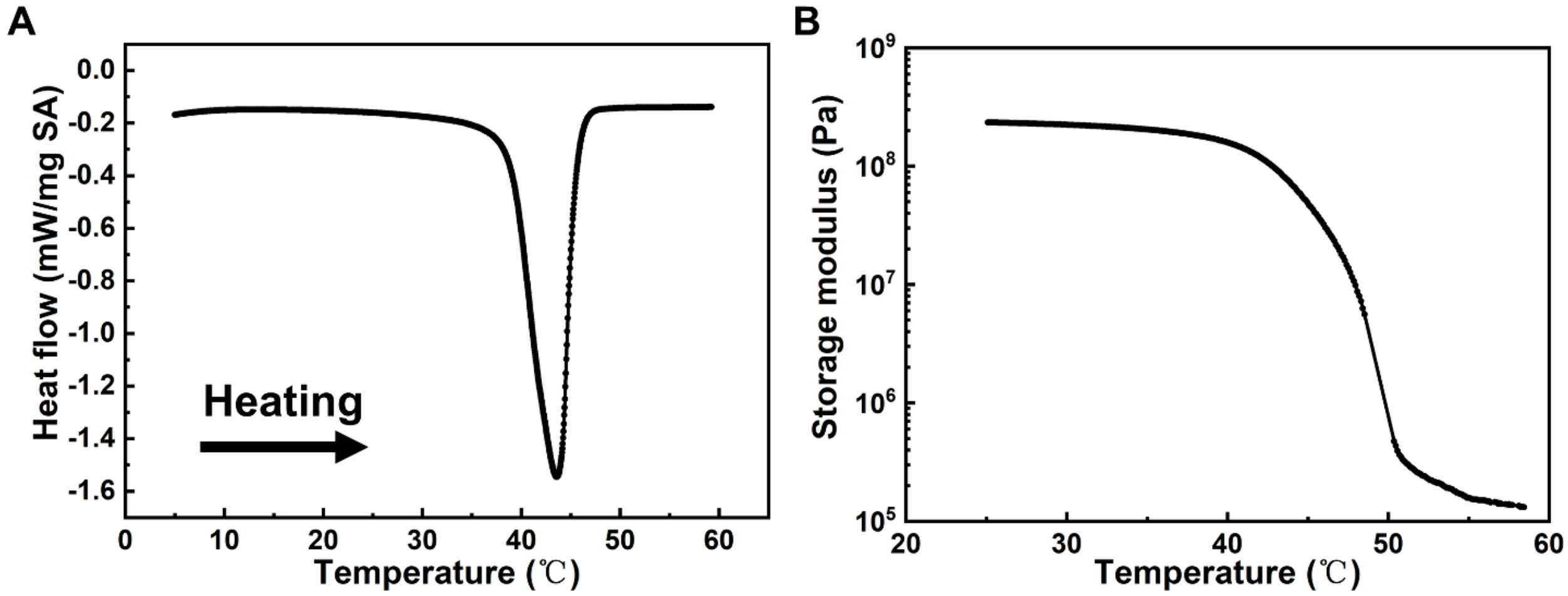

**Fig. S1. Thermal and mechanical properties of SPRR polymer. (A)** Heat flow measurement in DSC results of SPRR polymer during heating at a rate of 5 ℃/min. **(B)** Storage modulus measurement of SPRR polymer in DMA results as a function of temperature with a temperature ramping rate of 2 ℃/min.

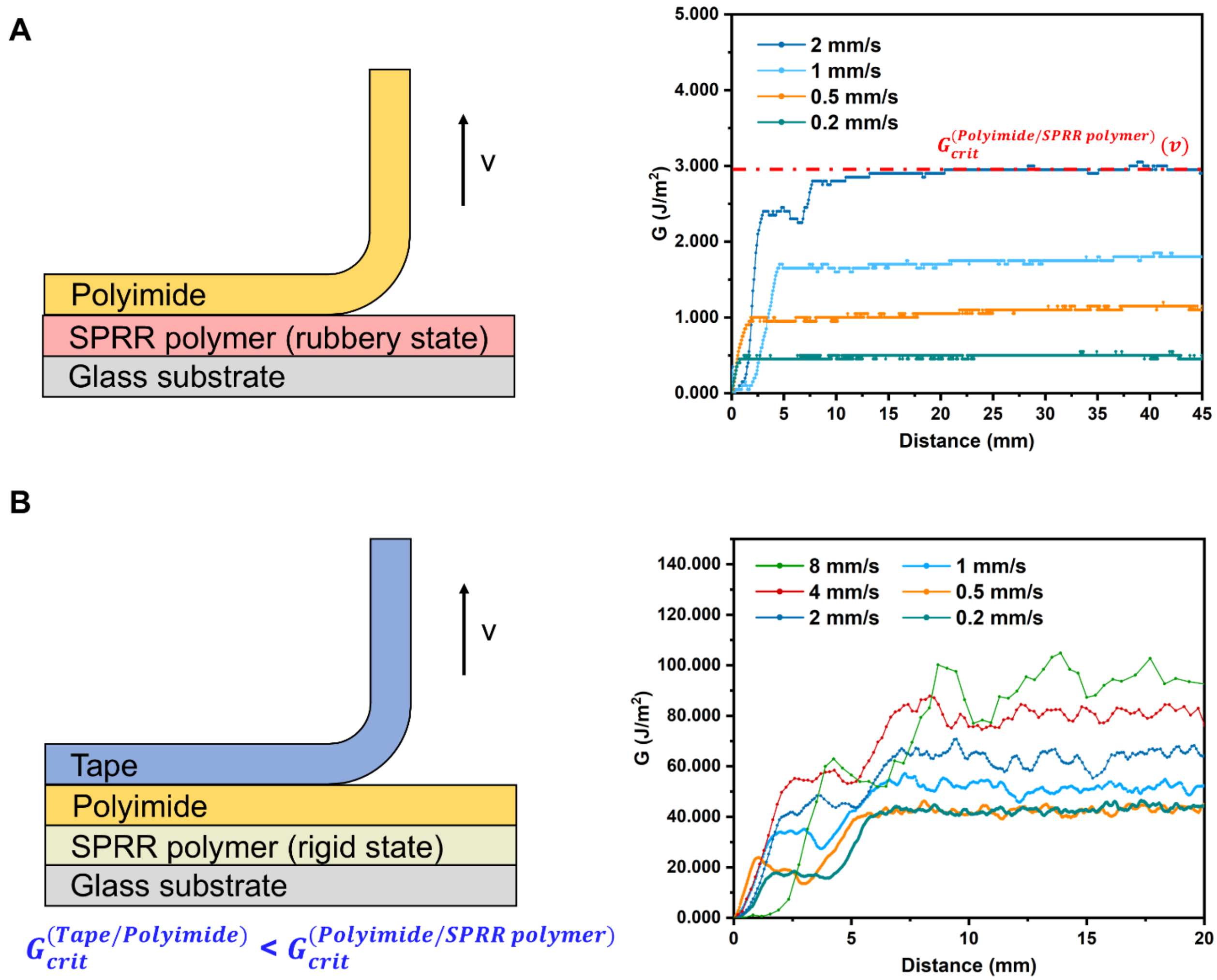

**Fig. S2. Adhesion characterization via 90-degree peeling test. (A)** Measurement of kinetically controlled adhesion between rubbery SPRR polymer and polyimide film. **(B)** Tape peeling experiments on polyimide film locked by rigid SPRR polymer.

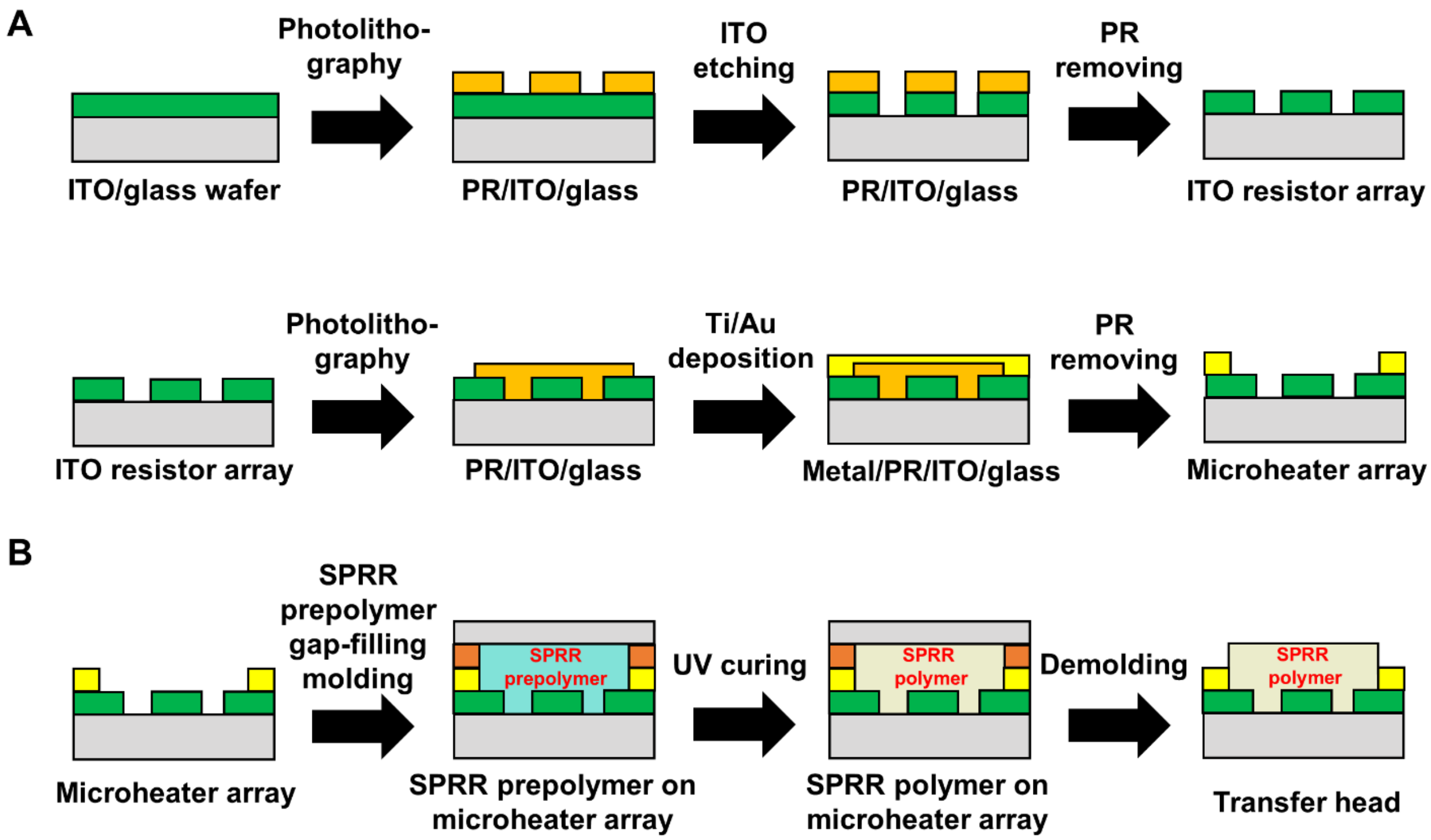

**Fig. S3. Fabrication process flow of the transfer head. (A)** Fabrication process of microheater. **(B)** SPRR polymer preparation on microheater.

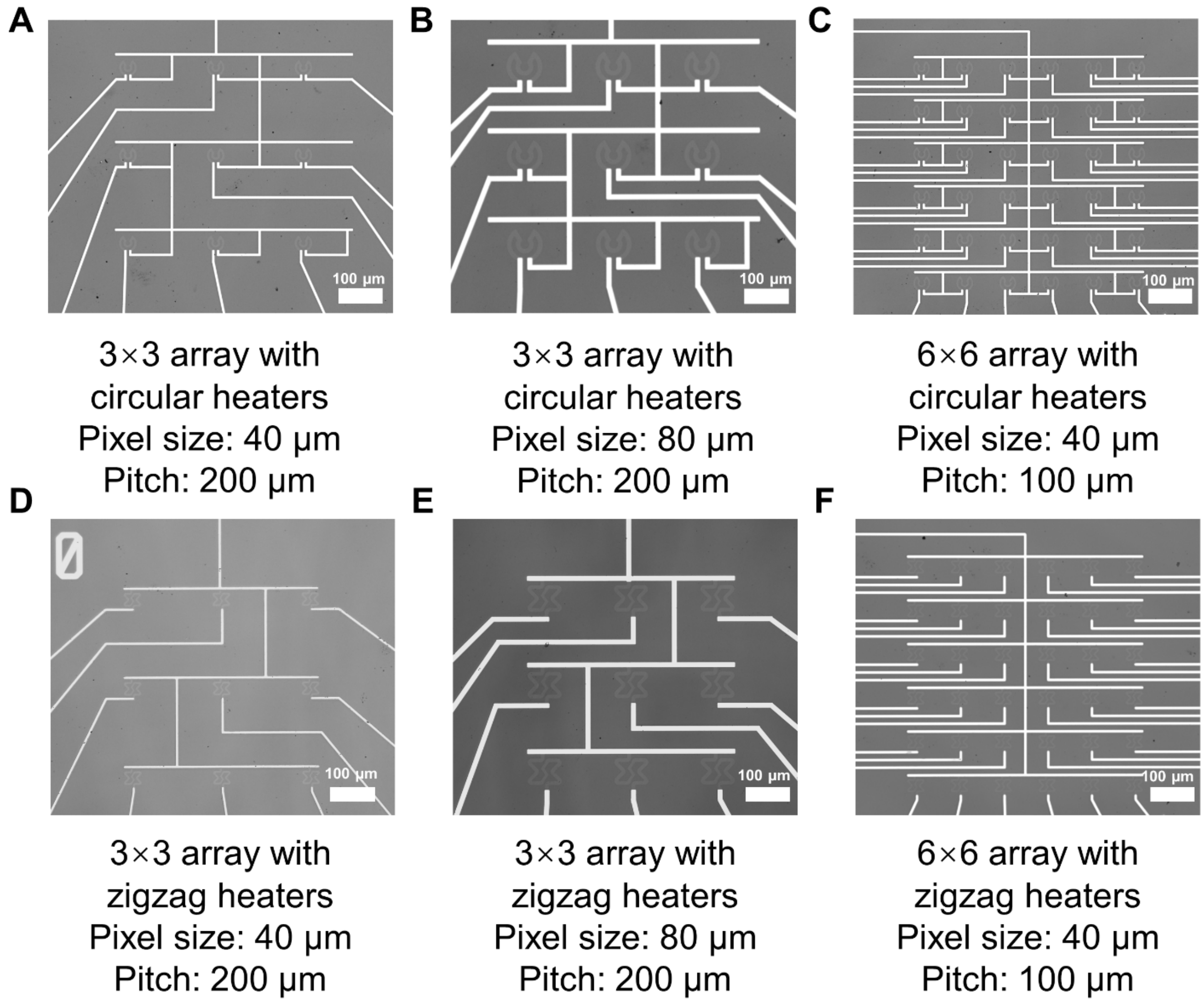

**Fig. S4. Multiple microheater arrays with different designs.** **(A)** 3 × 3 circular microheater array (Pixel size: 40 μm; Pitch: 200 μm). **(B)** 3 × 3 circular microheater array (Pixel size: 80 μm; Pitch: 200 μm). **(C)** 6 × 6 circular microheater array (Pixel size: 40 μm; Pitch: 100 μm). **(D)** 3 × 3 zigzag microheater array (Pixel size: 40 μm; Pitch: 200 μm). **(E)** 3 × 3 zigzag microheater array (Pixel size: 80 μm; Pitch: 200 μm). **(F)** 6 × 6 zigzag microheater array (Pixel size: 40 μm; Pitch: 100 μm).

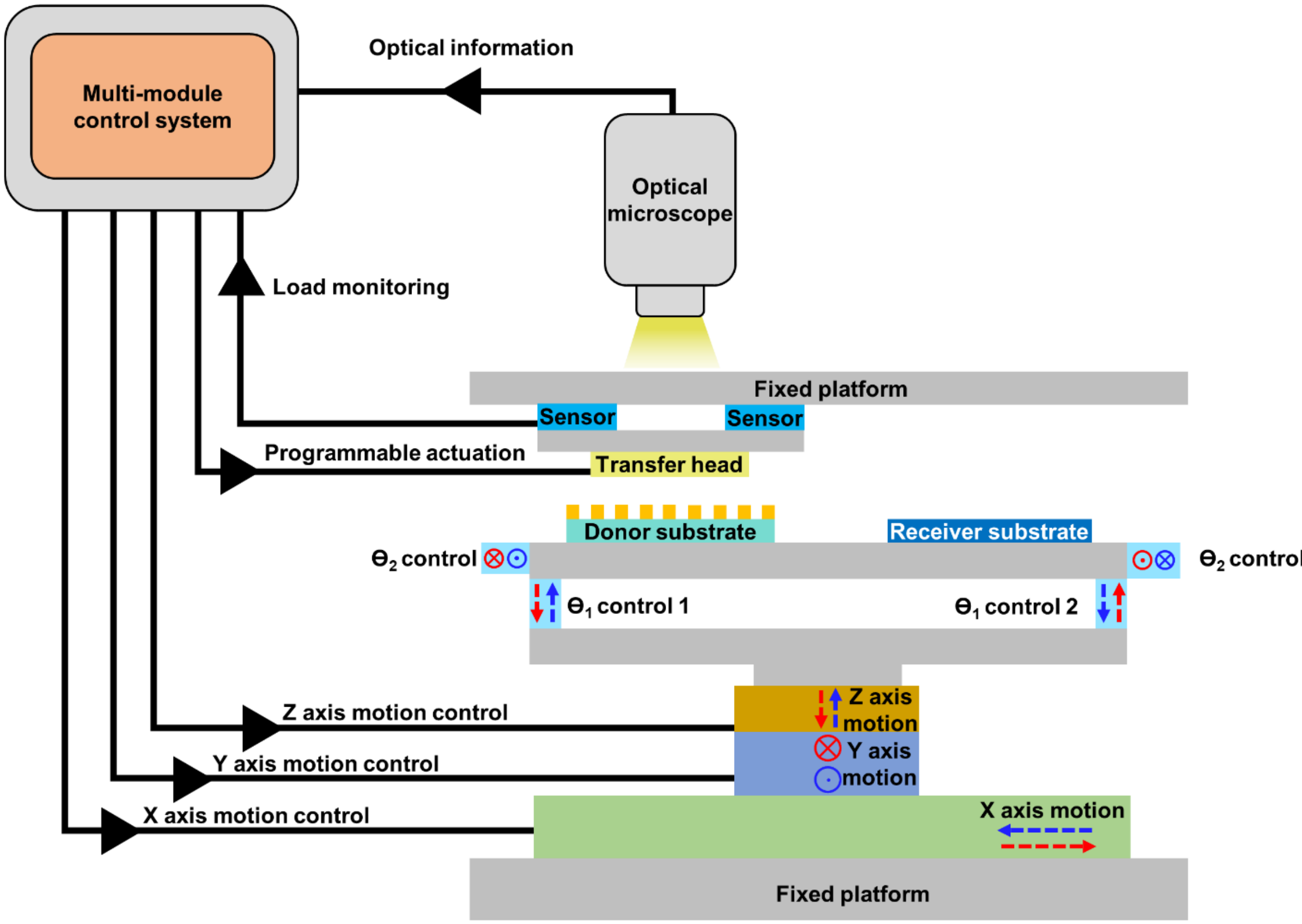

**Fig. S5. Schematic diagram of the high-precision dynamically programmable transfer printing system.**

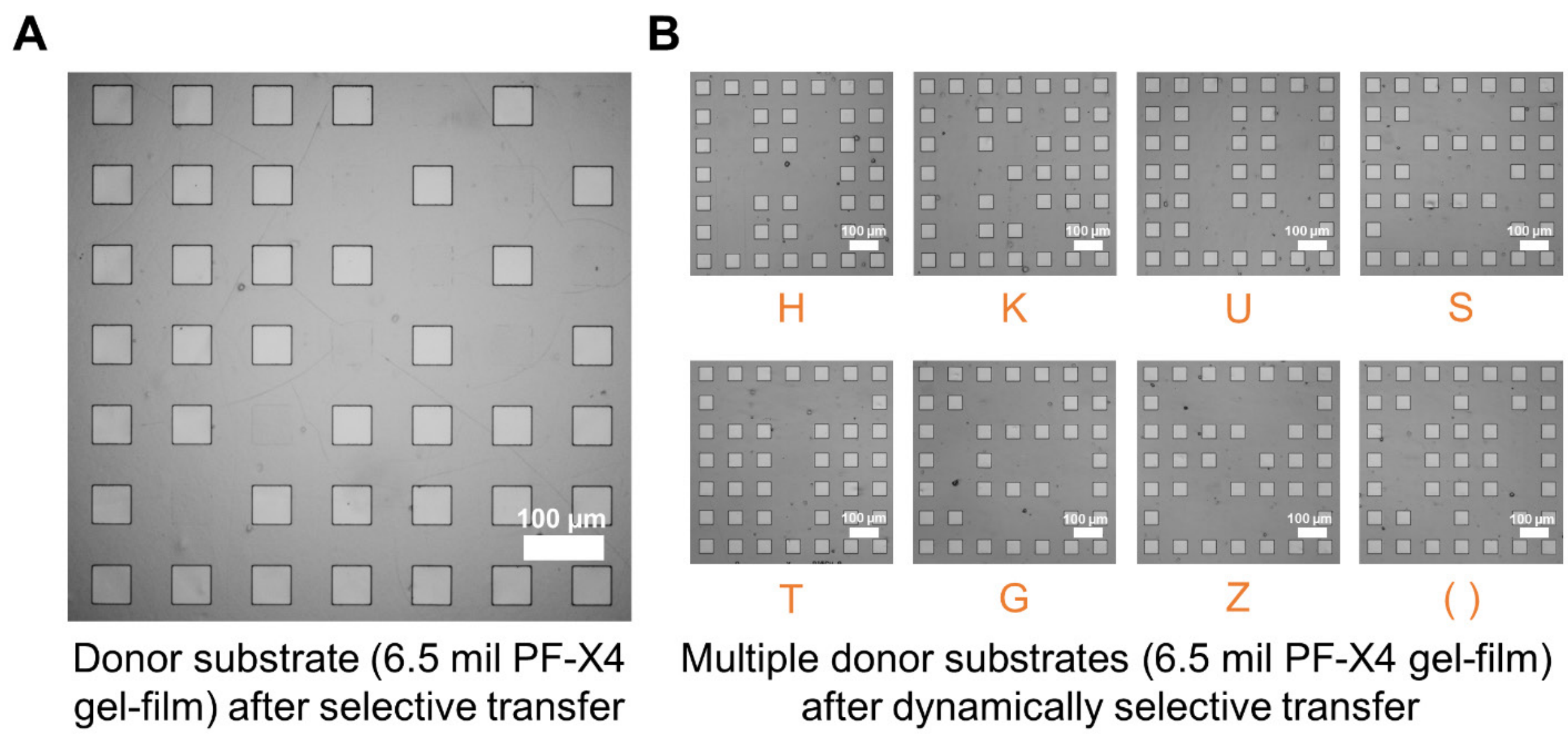

**Fig. S6. Multiple donor substrates with AZ5214E photoresist chiplets after dynamically programmable transfer. (A)** Donor substrate (6.5 mil thick PF-X4 gel-film) after selective transfer of AZ5214E photoresist chiplets. **(B)** Donor substrates (6.5 mil thick PF-X4 gel-film) after dynamically programmable transfers of AZ5214E photoresist chiplets for creating "HKUST(GZ)" pattern.

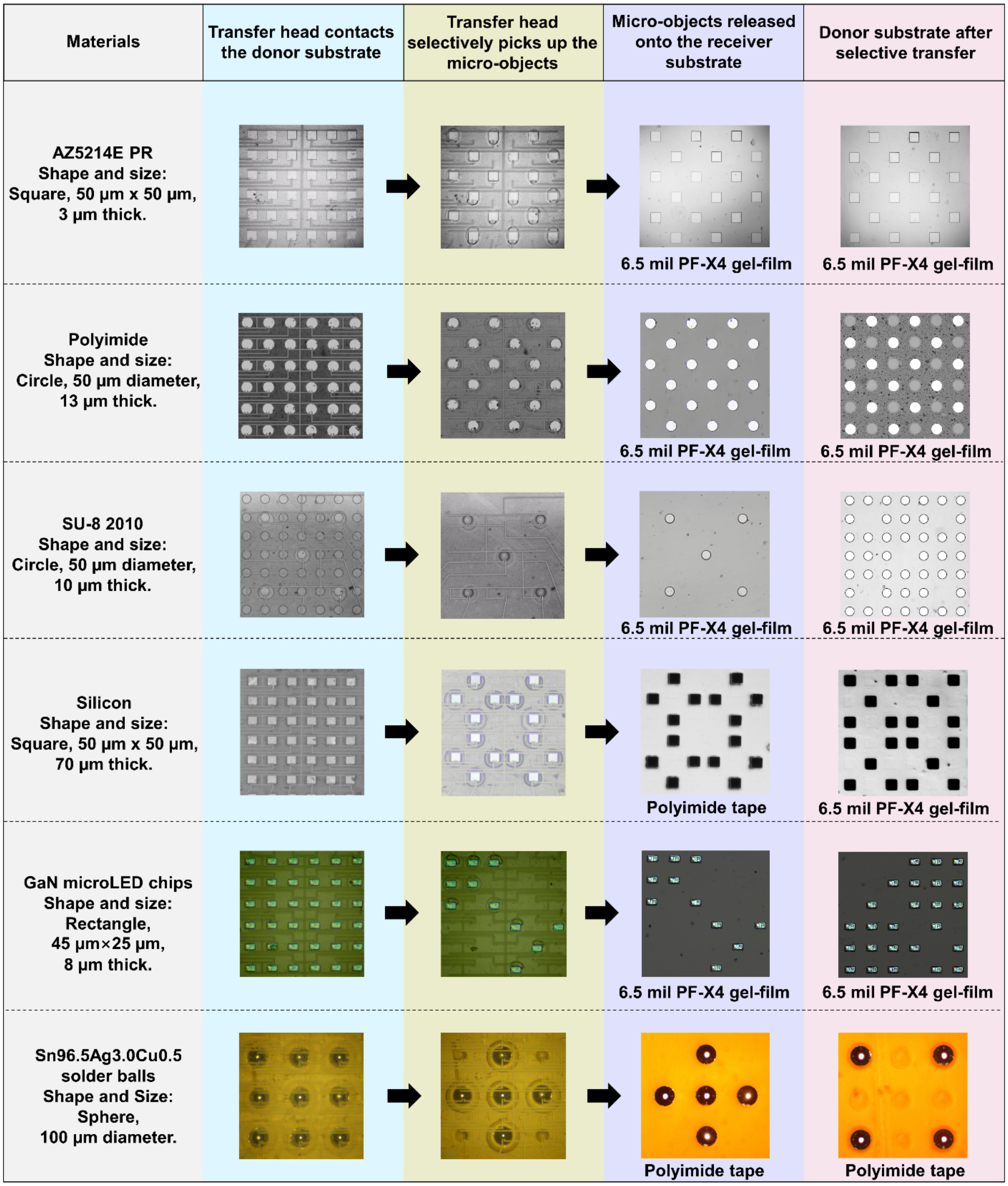

**Fig. S7. Multiple dynamically programmable transfer demonstrations of various materials ranging from microchiplets to microspheres.**

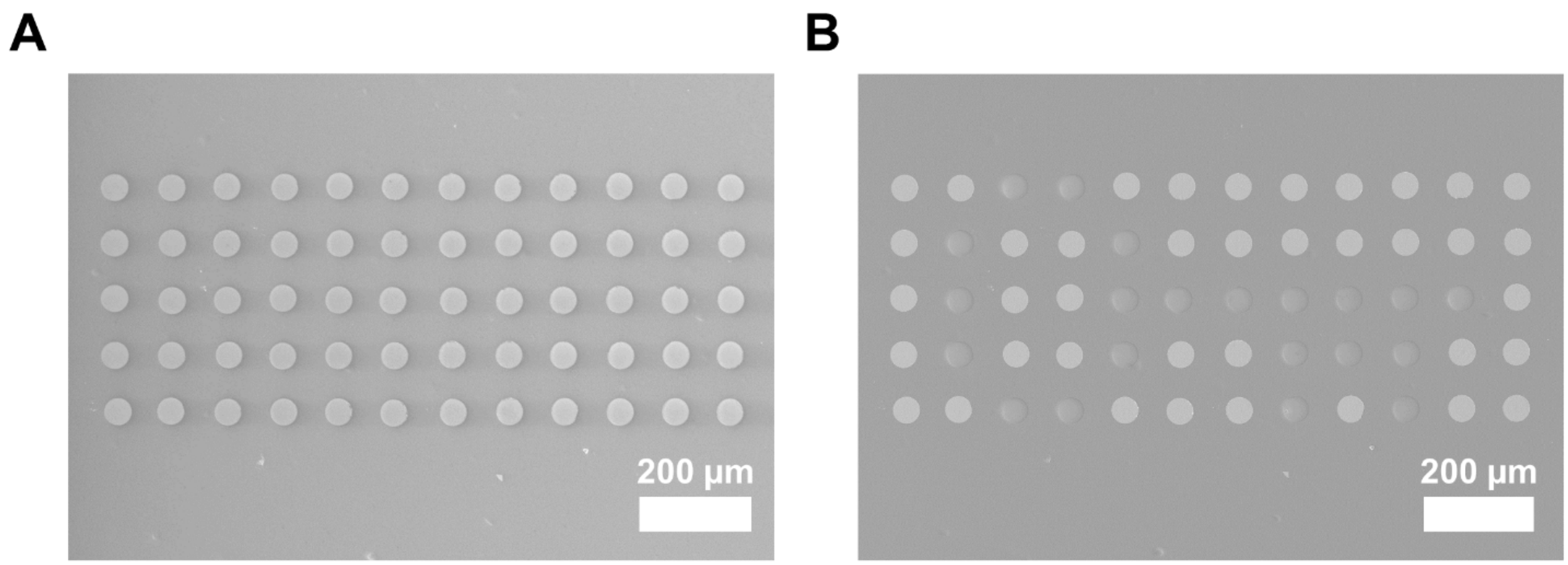

**Fig. S8. SEM images of 50-µm-diameter 90-nm thick copper film array on the donor substrate (6.5 mil thick PF-X4 gel-film) (A) before and (B)after dynamically programmable transfer.**

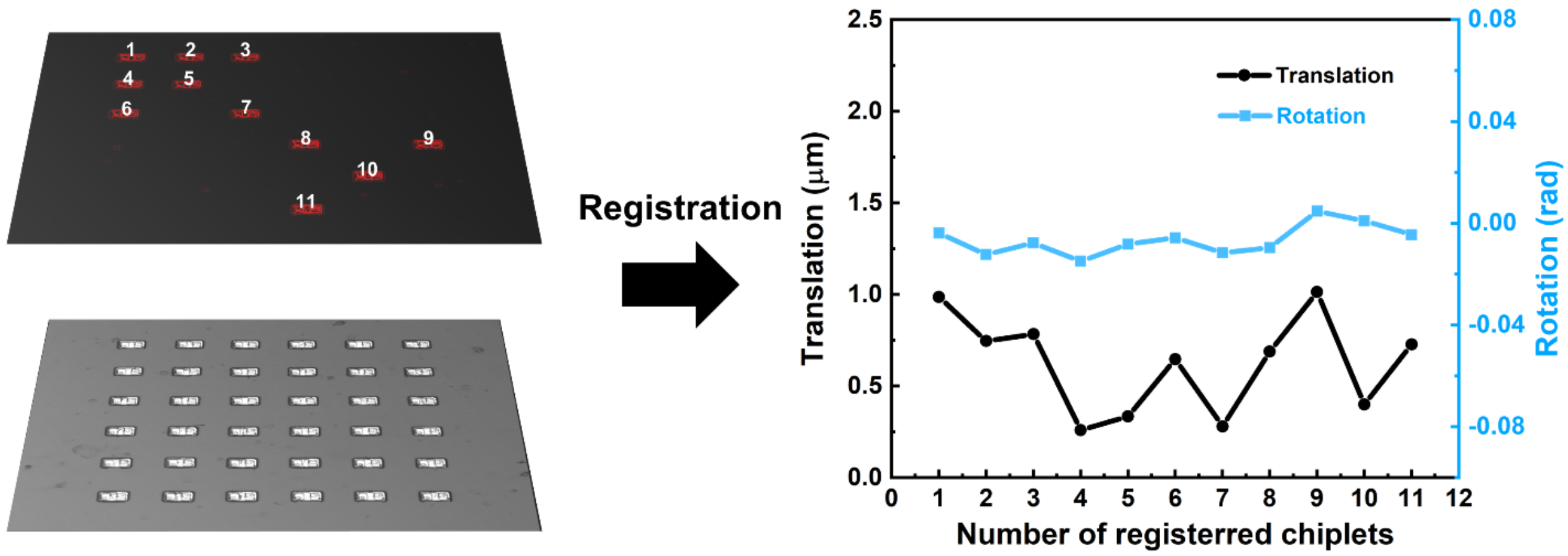

**Fig. S9. Registration error of microLED chips before and after dynamically programmable transfer.**

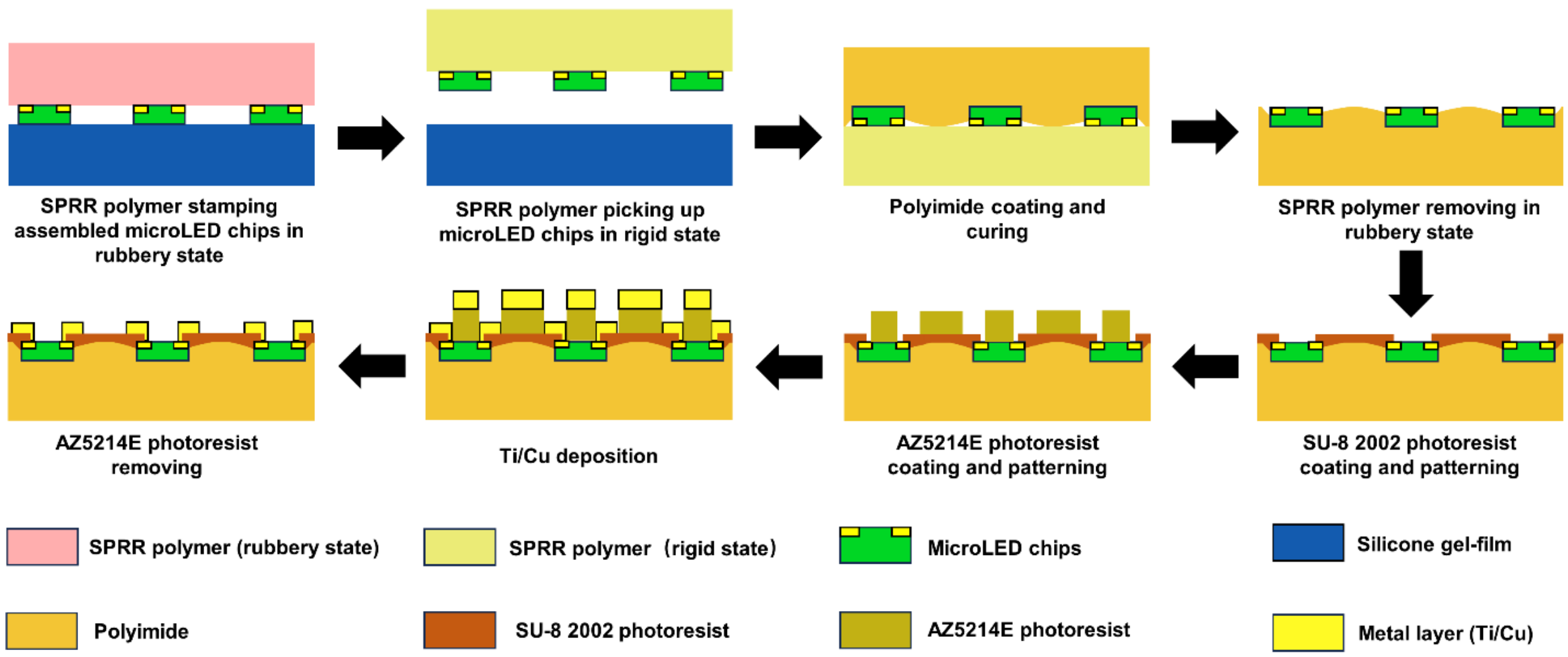

**Fig. S10. Process flow of flexible microLED signage fabrication.**

**Movie S1. Selective actuation of micro-transfer printing pixels.**

**Movie S2. Dynamically programmable transfer of 50 µm × 50 µm, 2-µm-thick AZ5214E photoresist chiplets.**

**Movie S3. Dynamically programmable transfer of different materials including polyimide microchiplets and polystyrene microspheres.**

**Movie S4. Flexible 3D stacking and heterogeneous integration.**

**Movie S5. Dynamically programmable transfer of microLED chips.**